\documentstyle[12pt,aps,prl,epsf]{revtex}

\begin{document}
\draft
\title{Dynamics of optically driven exciton and quantum decoherence}
\author{G.R. Jin, Yu-xi Liu$^{\dag}$, D.L. Zhou, X.X. Yi and C.P. Sun\cite{email,www}}
\address{Institute of Theoretical Physics,Academia Sinica,Beijing, 100080,China\\
$^{\dag}$The Graduate University for Advanced Studies (SOKEN),\\
Hayama, Kanagawa, 240-0193, Japan\\
\medskip}
\author{\parbox{14.2cm}{\small \hspace*{3mm}
By using the normal ordering method, we study the state evolution
of an optically driven excitons in a quantum well immersed in a
leaky cavity, which was introduced by Yu-xi Liu et.al. [Phys. Rev.
A {\bf 63}, 033816 (2001)]. The influence of the external laser
field on the quantum decoherence of a mesoscopically superposed
states of the excitons is investigated. Our result shows that,
thought the characteristic time of decoherence does not depend on
the external field, the phase of the decoherence factor can be
well controlled by adjusting the external parameters.\vskip30pt
PACS number(s): 42.50 Fx, 71.35-y}} \maketitle

\flushbottom \narrowtext \vskip20pt

\section{\ Introduction}

One of the most profound feature of quantum mechanics is the
superposition principle, which plays a central role in
implementing of quantum information processing (QIP)\cite{QI},
such as quantum computation, quantum cryptography and quantum
teleportation. The realization of QIP has triggered intense study
in various quantum system including ion traps, cavity QED, nuclear
magnetic resonant, and quantum dots.

As is well known, within the quantum information process
preserving coherence is an essential requirement. However, quantum
coherence is not robust enough to be exploited. In measurement
process the information's readout will lead to wave pocket
collapse, i.e., quantum decoherence. On the other hand, due to the
influence of the environment \cite{zur}, the system's coherence
information will be lost and the superposition of the system
states will evolve into statistical mixture state. In fact there
are two distinct effects of the external world on the quantum
system: quantum dissipation \cite{cal,yu&sun} with energy loss,
and quantum decoherence \cite{zur2} without energy loss. Both the
two effects will reduce the efficiency of quantum computation, and
result in disentanglement in QIP \cite{sun}. It is obvious that to
overcome the obstacle caused by decoherence and to control
decoherence now become more urgent.

Some important progresses have been made in the solid-state system
\cite{op,bayer} recently. The schemes to generate maximally
entangled states for excitons in coupled quantum dots have been
proposed by using a classical laser field \cite{quiroga} or a
quantum laser field \cite{yi}. Besides, a quantum superposition of
macroscopically distinct states \cite{QS} in a superconducting
quantum interference device (SQUID) has been demonstrated
experimently. These achievements have manifested that the
possibility of implement QIP in the solid system becomes more
promising than ever. In our previous works \cite{liu2}, the
quantum decoherence of a mesoscopically superposed state of the
excitons in a quantum well placed in a leaky cavity is
investigated. The results show that the coherence of the
superposed states of the system will undergo oscillating decay
with time evolution. Now, an immediately-followed question is that
how to control the dynamical evolution and to suppress the
decoherence of the system.

In this paper, with the motivation to continue our previous work,
we study the dynamical evolution of an optically driven exciton in
the quantum well placed in a leaky cavity. The effect of the
external continuous wave (c.w.) field on the state evolution and
the quantum decoherence of the mesoscopically superposed states of
the exciton is studied by using the normal ordering method (NOM)
\cite{louisell,jin}. Our result shows that the c.w. field does not
change the decoherence time, however, the phase of the decoherence
factor can be controlled by adjusting the amplitude of the field
and the detuning between the field and the transition frequency of
the two-level atoms. Such a result seems to be important in the
quantum computation because in a quantum computation the phase of
decoherence factor plays more crucial role \cite{sun}.

The paper is organized as follows: In section II, we firstly give
a model of the optically driven excitons in the quantum well
placed in a leaky cavity. In section III, the dynamical evolution
of the whole system is analytically calculated by directly solving
the Schr\"{o}dinger equation with the help of the normal ordering
method. In addition, the influence of the external pumping field
on the mean population of the exciton system is also studied. In
section IV, we study the decoherence behavior of the optically
driven exciton system. The effect of the external field on the
decoherence process of the exciton is investigated carefully.
Finally, we will give some conclusions.

\section{ Model of an optically driven exciton in a leaky cavity}
In our previous work \cite{liu2,liu1}, we considered a quantum
well (or crystal slab) placed within a leaky Fabry-Perot
cavity~\cite{scu}. The quantum well lies in the center of the
cavity. It contains an ideal cubic lattice with $N$ lattice sites
and is so thin that it has only one layer. We assume that $N$
identical lattice two-level atoms (or molecules) distribute into
these lattice sites. All these particles have equivalent mode
positions, so they have the same coupling constant with the cavity
modes. It is also assumed that the direction of the dipole moment
for the two-level atoms and wave vectors of the cavity fields are
perpendicular to the surface of the slab. In addition, when a
continuous wave (c.w.) pumping field with frequency $\omega $ is
applied on the quantum well, the total Hamiltonian under the
rotating wave approximation becomes, with $\hbar =1,$

\begin{equation}
H=\Omega S_z+\sum_j\omega _jb_j^{\dagger
}b_j+\sum_jg_j(b_jS_{+}+b_j^{\dagger }S_{-})+\Re \left( e^{-i\omega
t}S_{+}+e^{i\omega t}S_{-}\right)
\end{equation}
with the collective operators
\begin{equation}
S_Z=\sum_{n=1}^Ns_z(n),\quad S_{\pm }=\sum_{n=1}^Ns_{\pm }(n),
\end{equation}
where $s_z(n)=\frac 12(|e_n\rangle \langle e_n|-|g_n\rangle \langle g_n|)$, $%
s_{+}(n)=|e_n\rangle \langle g_n|$ and $s_{-}(n)=|g_n\rangle \langle e_n|$
are quasi-spin operators of the n-th atom. Here $|e_n\rangle $ and $%
|g_n\rangle $ denote the excited state and the ground state of n-th atom,
and $\Omega $ is a transition frequency of the isolated atom. Operators $%
b_j^{\dagger }(b_j)$ are creation (annihilation) operators of the field
modes which labeled by continuous index $j$ with mode frequency $\omega _j$.
The Rabi frequency $\Re $ denotes the coupling between the atoms and the
c.w. pumping field. The coupling constant $g_j$ between the molecules and
the cavity fields takes a simple form which is proportional to a Lorentzian
\begin{equation}
g_j=\frac{\eta \Gamma }{\sqrt{(\omega _j-\Omega )^2+\Gamma ^2}},
\end{equation}
where $\eta $ depends on the atomic dipole~and $\Gamma $ is the
decay rate of a quasi-mode of the cavity. In this paper we
restrict our investigation to the Jaynes-Cummings situation where
only one quasi-mode of the cavity is
involved and is resonant with the transition frequency of the isolate atom $%
\Omega $ \cite{Law}. In the case of low density of the excitation,
with the attractive exciton-exciton collisions due to the
bi-exciton effect \cite {liu1} neglected, the collective behavior
of the atoms can be described by a bosonic exciton~\cite{Hak}.
With this so-called bosonic approximation we can make
the replacement: $a=S_{-}/\sqrt{N}$ and $a^{\dagger }=S_{+}/\sqrt{N}$ with $%
[a,a^{\dagger }]=1$. Then the Hamiltonian (1) becomes

\begin{equation}
H=\Omega a^{\dagger }a+\sum_j\omega _jb_j^{\dagger }b_j+\sum_jg(\omega
_j)(b_j^{\dagger }a+a^{\dagger }b_j)+\xi \left( e^{-i\omega t}a^{\dagger
}+e^{i\omega t}a\right) ,
\end{equation}
with $g(\omega _j)=\sqrt{N}g_j$ and $\xi =\sqrt{N}\Re $. We note
that our model of the optically driven excitons plus the cavity
fields now becomes a standard driven damped oscillator system. We
will solve the time evolution of the coupled system described by
the Hamiltonian of eq.(4) by directly solving the Schr\"{o}dinger
equation with the help of the NOM.

\section{Exact solution in terms of NOM}

In this section we calculate the state evolution of the driven
excitons immersed in a lossy cavity by using NOM, which is firstly
introduced by Louisell \cite{louisell} to study the dynamic
evolution of a driven oscillator as well as that of two weakly
coupled oscillators without dissipation. Here we follow our
previous works \cite{jin} to study the dynamical evolution of the
optically driven exciton immersed in a leaky cavity.

The state vector of the whole system obeys the Schr\"{o}dinger equation
which has a solution of the form $\left| \psi (t)\right\rangle =U(t)\left|
\psi (0)\right\rangle $, where the time-evolution operator $U(t)$ satisfies $%
i\partial _tU(t)=HU(t),$with the initial condition $U(0)=1$. We assume that
evolution operator has its normal order form $U(t)=U^{(n)}(t)$. Because the
normal form of any operator is unique, one can establish the one-to-one
corresponding relationship between the normal ordered evolution operator $%
U^{(n)}(t)$ and an ordinary function $\overline{U}^{(n)}(t)$, with $%
\overline{U}^{(n)}(t)=\left\langle \alpha ,\left\{ \beta _j\right\} \right|
U^{(n)}(t)\left| \left\{ \beta _j\right\} ,\alpha \right\rangle $. Here $%
\left| \left\{ \beta _j\right\} \right\rangle =\prod_j\left| \beta
_j\right\rangle $ denotes multimode coherent state of the radiation field.
Such a corresponding relation defines a map $\aleph ^{-1}$ ,

\begin{equation}
\aleph ^{-1}:U^{(n)}(t)\rightarrow \overline{U}^{(n)}(t).
\end{equation}
On the other hand, we can also define the inverse transformation
$\aleph $ ,

\begin{equation}
\aleph :\overline{U}^{(n)}(t)\rightarrow U^{(n)}(t)=U(t).
\end{equation}
Therfore, one can write down the Schr\"{o}dinger equation of
$U(t)$ in the normal ordering form. Then implementing the operator
$\aleph ^{-1}$, one can get a c-number equation of
$\overline{U}^{(n)}$,
\begin{eqnarray}
i\partial _t\overline{U}^{(n)} &=&[\Omega \alpha ^{*}\left( \alpha +\partial
_{\alpha ^{*}}\right) +\sum_j\omega _j\beta _j^{*}(\beta _j+\partial _{\beta
_j^{*}})+\sum_jg(\omega _j)\beta _j^{*}(\alpha +\partial _{\alpha ^{*}})
\nonumber \\
&&+\sum_jg(\omega _j)\alpha ^{*}(\beta _j+\partial _{\beta _j^{*}})+\xi
e^{-i\omega t}\alpha ^{*}+\xi e^{i\omega t}\left( \alpha +\partial _{\alpha
^{*}}\right) ]\overline{U}^{(n)}.
\end{eqnarray}
where $\partial _{\alpha ^{*}}=\frac \partial {\partial \alpha ^{*}}$, $%
\partial _{\beta _j^{*}}=\frac \partial {\partial \beta _j^{*}}$.

We assume $\overline{U}^{(n)}$ takes the form of

\begin{eqnarray}
\overline{U}^{(n)} &=&\exp \{A+B\alpha +C\alpha ^{*}+D\alpha ^{*}\alpha
+\sum_jB_j\beta _j^{*}\beta _j+{\sum_{j,j^{\prime }}}^{\prime
}B_{j,j^{^{\prime }}}\beta _j^{*}\beta _{j^{^{\prime }}}  \nonumber \\
&&+\sum_jC_j\beta _j^{*}\alpha +\sum_jD_j\alpha ^{*}\beta _j+\sum_jE_j\beta
_j^{*}+\sum_jF_j\beta _j\}.
\end{eqnarray}
Here, the prime in ${\sum }^{\prime }$ denotes sum over index $``j"$ and $%
``j^{^{\prime }}"$ with the condition $j\neq j^{^{\prime }}$. By
substituting eq.(8) into eq.(7), we get equations which the
time-dependent coefficients obeys (for the details of
calculations, please see Appendix). In this paper we restrict our
study to zero temperature situation for the fields where no
background radiation is involved in our consideration. At zero
temperature the radiation field of the present model is in its
pure vacuum state $\left| \{0_j\}\right\rangle =\prod_j\left|
0_j\right\rangle $. Thus only the coefficients: $A$, $B$, $C$,
$D$, $C_j$\ and $E_j$\ contribute to the state evolution of the
whole system. We can write down the explicit expressions of these
coefficients as,

\begin{equation}
D(t)=u(t)e^{-i\Omega t}-1,  \eqnum{9a}
\end{equation}

\begin{equation}
C(t)=B(t)e^{-i\omega t}=w(t)e^{-i\Omega t},  \eqnum{9b}
\end{equation}

\begin{equation}
A(t)=-i\xi \int_0^tdt^{^{\prime }}w(t^{^{\prime }})e^{i\delta t^{^{\prime
}}},  \eqnum{9c}
\end{equation}

\begin{equation}
C_j(t)=u_j(t)e^{-i\Omega t},  \eqnum{9d}
\end{equation}

\begin{equation}
E_j(t)=v_j(t)e^{-i\Omega t},  \eqnum{9e}
\end{equation}
for
\begin{equation}
u(t)=[cos(\Theta t)+\frac \Gamma {2\Theta }sin(\Theta t)]e^{-\frac \Gamma 2%
t},  \eqnum{10}
\end{equation}
with ${\Theta }=\sqrt{M\Gamma -(\Gamma /2)^2}$,

\begin{eqnarray}
u_j(t) &=&-\frac{g(\omega _j)}2(1-\frac{i\Gamma }{2\Theta
})\frac{e^{i\Theta t}e^{-\frac \Gamma 2t}-e^{-i(\omega _j-\Omega
)t}}{\omega _j-\Omega +\Theta +i\frac \Gamma 2}  \nonumber \\
&&-\frac{g(\omega _j)}2(1+\frac{i\Gamma }{2\Theta })\frac{e^{-i\Theta t}e^{-%
\frac \Gamma 2t}-e^{-i(\omega _j-\Omega )t}}{\omega _j-\Omega -\Theta +i%
\frac \Gamma 2},  \eqnum{11}
\end{eqnarray}
and the coefficient $w(t)$ is

\begin{eqnarray}
w(t) &=&-\frac \xi 2(1-\frac{i\Gamma }{2\Theta })\frac{e^{i\Theta t}e^{-%
\frac \Gamma 2t}-e^{-i\delta t}}{\delta +\Theta +i\frac \Gamma 2}
\nonumber
\\
&&-\frac \xi 2(1+\frac{i\Gamma }{2\Theta })\frac{e^{-i\Theta t}e^{-\frac %
\Gamma 2t}-e^{-i\delta t}}{\delta -\Theta +i\frac \Gamma 2}.
\eqnum{12}
\end{eqnarray}
 where $\delta =\omega -\Omega $ is the detuning between the
c.w. field with the transition frequency of the two-level atom.
The coefficients $A(t)$, and $v_j(t)$\ can also be solved. Here we
do not give the explicit expressions for brevity.

If, as an example, the initial state of the total system is
$\left| \psi (0)\right\rangle =\left| \alpha \right\rangle \otimes
\left| \left\{ 0_j\right\} \right\rangle$. Then at any time $t$
the total system will evolve into

\begin{equation}
\left| \psi (t)\right\rangle =\left| \alpha u(t)e^{-i\Omega
t}+w(t)e^{-i\Omega t}\right\rangle \otimes \left| \left\{ \alpha
u_j(t)e^{-i\Omega t}+v_j(t)e^{-i\Omega t}\right\} \right\rangle ,  \eqnum{13}
\end{equation}
from which we can determine the mean number of excitons at time $t$ as

\begin{equation}
\bar{n}(t)=\bar{n}_0|u(t)|^2+|w(t)|^2+\alpha u(t)w^{*}(t)+c.c.,  \eqnum{14}
\end{equation}
where $\bar{n}_0=|\alpha |^2$ is the initial mean number of
excitons. In Fig.1, we give a sketch of the mean number of
excitons as the function of time. We find that the c.w. field
compensates the loss of population of excitons due to the damping
of cavity \cite{oliver}. By adjusting the detuning between the
external field with the transition frequency of the two-level
atoms and the amplitude of the field we can obtain different value
of mean number of excitons at long time limit.

\vskip 0.2cm
\begin{figure}
\epsfxsize=10cm\epsffile{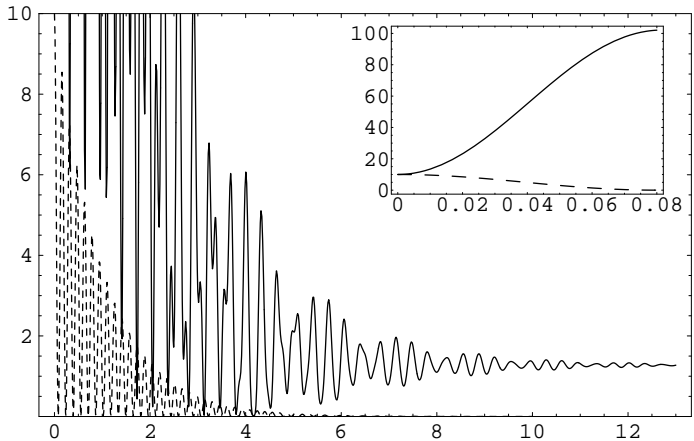} \caption{The mean number of
excitons as the function of time with a given set of parameters:
$\hbar \Gamma=0.05$ meV, $\bar{n}_0=10$, $\hbar M=20$meV. No
pumping is applied (dashed curves), $\hbar \xi=10$meV, $\hbar
\delta=0.1$ meV (solid curves). The inset shows the short time
behavior of mean number of excitons with the same set of the above
parameters.}
\end{figure}

\section{Quantum decoherence of the optically driven exciton system}

If we consider a superposition of distinct coherent states as
Schr\"{o}dinger's cat, i.e., the excitons is initially in the state $%
C_1\left| \alpha _1\right\rangle +C_2\left| \alpha _2\right\rangle $, where $%
\left| \alpha _1\right\rangle$ and $\left| \alpha _2\right\rangle$
are coherent states of the exciton, and the cavity fields are in
the vacuum states $\left| \{0_j\}\right\rangle$. Then the state
vector at any time $t$ is

\begin{eqnarray}
|\Psi (t)\rangle =C_1\exp \{(A+B\alpha _1)/2-c.c.\}|\alpha
_1ue^{-i\Omega t}+we^{-i\Omega t}\rangle  &&\otimes |\left\{
\alpha _1u_je^{-i\Omega t}+v_je^{-i\Omega t}\right\} \rangle
\nonumber \\ +C_2\exp \{(A+B\alpha _2)/2-c.c.\}|\alpha
_2ue^{-i\Omega t}+we^{-i\Omega t}\rangle  &&\otimes |\left\{
\alpha _2u_je^{-i\Omega t}+v_je^{-i\Omega t}\right\} \rangle,
\eqnum{15}
\end{eqnarray}
where we have used the following "sum rules",

\begin{eqnarray}
|\alpha _1|^2=\sum_j\left| \alpha _1u_j+v_j\right| ^2+\left|
\alpha _1u+w\right| ^2 &&+(A+B\alpha _1+c.c.),  \eqnum{16a} \\
|\alpha _2|^2=\sum_j\left| \alpha _2u_j+v_j\right| ^2+\left|
\alpha _2u+w\right| ^2 &&+(A+B\alpha _2+c.c.),  \eqnum{16b}
\end{eqnarray}
and,

\begin{eqnarray}
\langle \alpha _1|\alpha _2\rangle  &=&\exp \left\{ -(B\alpha
_1-c.c.)/2\right\} \exp \left\{ (B\alpha _2-c.c.)/2\right\}
\nonumber \\ &&\times \langle \{\alpha _1u_je^{-i\Omega
t}+v_je^{-i\Omega t}\}|\alpha _2u_je^{-i\Omega t}+v_je^{-i\Omega
t}\rangle   \nonumber \\ &&\times \langle \alpha _1ue^{-i\Omega
t}+we^{-i\Omega t}|\alpha _2ue^{-i\Omega t}+we^{-i\Omega t}\rangle
,  \eqnum{16c}
\end{eqnarray}
under the consideration of normalization condition of the wave
function. Eq. (15) is the main result of our study, from which we
see that, due to the fields fluctuation and the back-action of
system on the fields, the state vector evolved from factorized
initial state becomes fully entangled. However under certain
condition the total state vector can be partially factorized
\cite{Sun&Gao}. In the following context of this paper we will
investigate the effect of the external c.w. field on the quantum
decoherence of the superposition of the excitons.

We can calculate the reduced density matrix of the exciton system
$\rho (t)=Tr_R\left\{ |\Psi (t)\rangle \left\langle \Psi
(t)\right| \right\}$. Substituting eq.(15) into $\rho (t)$, we get
the decoherence factor, which is define as the coefficient of the
off-diagonal element of the reduced density matrix,
\begin{equation}
F(t)=\exp [(-\frac 12|\alpha _1|^2-\frac 12|\alpha _2|^2+\alpha
_1^{*}\alpha _2)(1-|u|^2)]e^{\frac 12(\alpha _1-\alpha
_2)uw^{*}-c.c.},  \eqnum{17}
\end{equation}
where we have used eq.(16c) in deriving eq.(17). The explicit
expressions of the time-dependent functions $u(t)$, and $
w(t)$ are given in eq.(10), and eq.(12), respectively. We consider that $%
\alpha _1=\alpha $, and $\alpha _2=\alpha e^{i\Delta \varphi }$, where $%
\Delta \varphi $ is the phase shift of the initial superposed states. The
characteristic time $\tau _d$ of the decoherence of the superposition state
is determined by the short time behavior of $|F(t)|$, that is $\Gamma
t,\Theta t\ll 1$. Within this time scale the norm of decoherence factor can
be simplified as following,

\begin{equation}
|F(t)|=\exp [-2|\alpha |^2\sin ^2\left( \Delta \varphi /2\right)
\Gamma t]. \eqnum{18}
\end{equation}
Then the characteristic time is determined as following,

\begin{equation}
\tau _d^{-1}=2|\alpha |^2\Gamma \sin ^2\left( \Delta \varphi
/2\right) , \eqnum{19}
\end{equation}
where $|\alpha |^2$ is the mean number of the excitons. We can define the
``distance'' $D=|\alpha _1-\alpha _2|=2|\alpha |\sin \left( \Delta \varphi
/2\right) $ between the two superposed states of the exciton. Substituting $D
$ into eq.(19), we get the characteristic time of decoherence $\tau _d=\frac{%
2\tau _p}{D^2}$, where $\tau _p=1/\Gamma $ is the life time of the
quasimode. Our result shows that the decoherence time of the exciton is
determined by the distance of their initial superposed states and the decay
rate of the quasimode. The external c.w. laser field dose not change the
decoherence time of the exciton. One can suppress the decoherence speed of
the exciton by adjusting the distance of the initial superposition of the
exciton \cite{liu2}.

For a special case $\Delta \varphi =\pi $, i.e., when the system
is prepared initially in odd or even coherent states
\cite{oliver,zoller,piza,c4} of the exciton. Then the decoherence
factor is

\begin{equation}
F(t)=\exp [-|\alpha |^2(1-|u|^2)]e^{i\phi (t)},  \eqnum{20}
\end{equation}
where we have introduced the phase of the decoherence factor in
eq.(20),

\begin{equation}
\phi (t)=%
\mathop{\rm Im}%
\left\{ \alpha u(t)w(t)^{*}-c.c.\right\} .  \eqnum{21}
\end{equation}
In Fig. 2, we plot the phase of the decoherence factor as the time
of $t$. We find that the phase depends linearly upon the amplitude
of the c.w. field (Fig. 2a). Besides, by adjusting the detuning
(Fig. 2b, and Fig. 2c) between the external field with the
transition frequency of the two-level atoms the phase factor can
also be controlled.

\vskip 0.2cm
\begin{figure}
\begin{tabbing}
\epsfxsize=10cm\epsffile{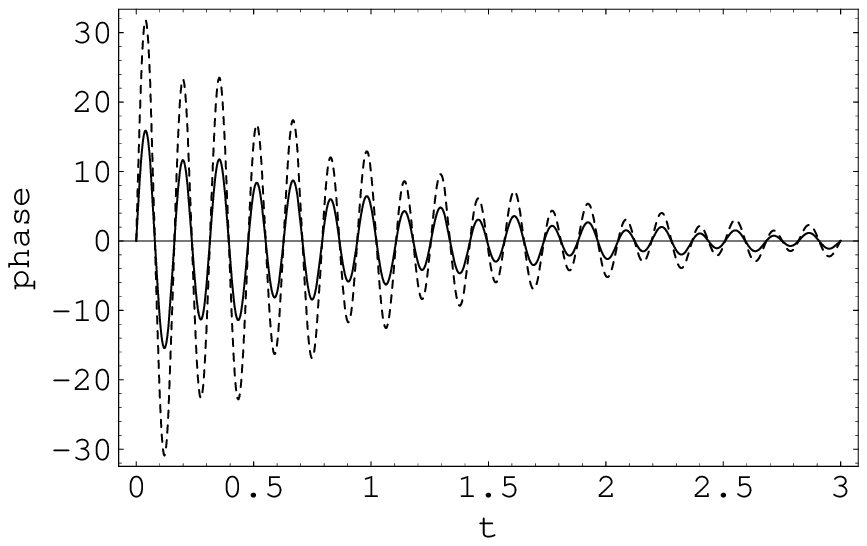}\\
\epsfxsize=10cm\epsffile{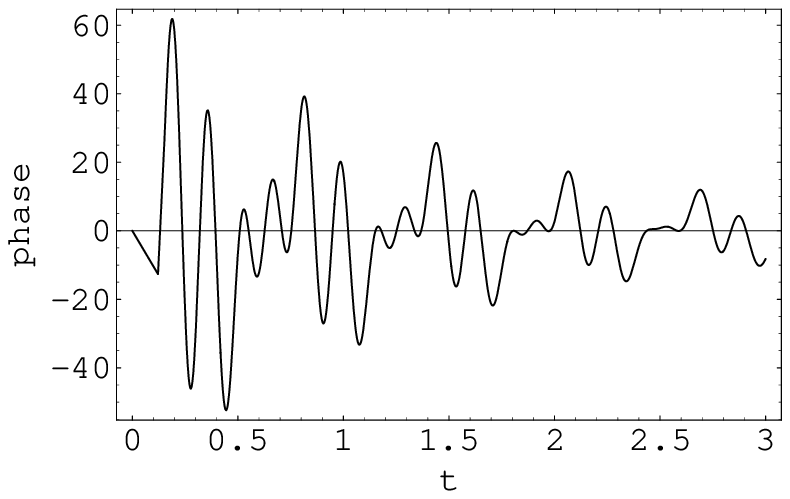}\\
\epsfxsize=10cm\epsffile{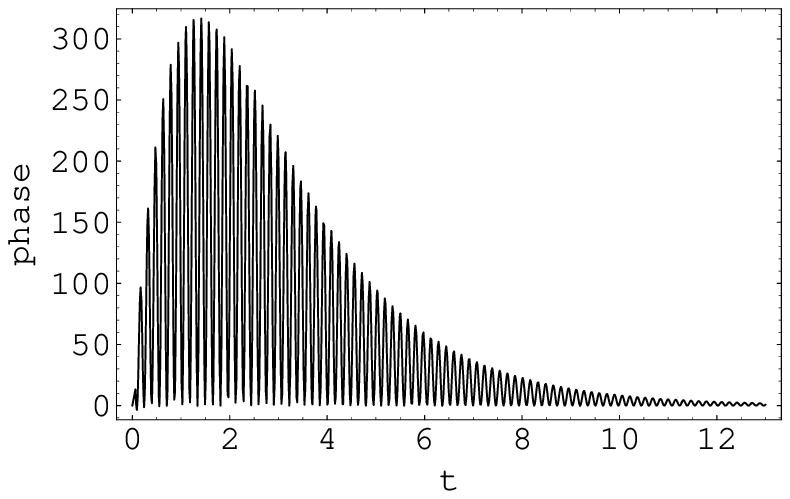}
\end{tabbing}
\vskip 0.2cm
\caption{the phase of the decoherence factor as the
time of $t$. $\hbar \Gamma =20$meV, $\hbar M=20$meV, and
$\bar{n}_0=10$, respectively. (a)$\hbar \delta =0$, $\hbar \xi
=10$meV (dashed curves), and $\hbar \xi =5$meV (solid
curves), (b)$\hbar \delta =0.5$meV, $\hbar \xi =10$meV, (c)$\hbar \delta =1$%
meV, $\hbar \xi =10$meV.}
\end{figure}

\section{Conclusions}

We have studied the dynamical evolution of superposition of
mesoscopically distinct quantum state in a system of an
optically-driven exciton in a quasimode cavity. By utilizing
normal ordering technique, the explicit expression of state vector
at any time is obtained in the case of no background radiation.
The influence of c.w. field on the mean number of excitons in the
lossy cavity is also studied. We find that the field compensates
the loss of population of excitons. By adjusting the external
parameters we can obtain different value of mean number of
excitons at long time limit.

By solving the explicit form of the decoherence factor, we
investigate the decoherence behavior of the exciton system and
found that the decoherence rate of the exciton does not depend on
the c.w. laser field. However the phase of the decoherence rate
can be well controlled by adjusting the amplitude of the external
field as well as the detuning between the field and the transition
frequency of the atom.

\acknowledgements This work is supported in part by the National Foundation
of Natural Science of China. One of authors (G.R. Jin) is indebted to
Professor He-shan Song for continuous encouragement in his work.

\appendix

\section*{Normal ordering method}

In this appendix, we will give the equations of the time-dependent
coefficients and the formal solution of the equations. By substituting
eq.(8) into eq.(7), we have
\begin{equation}
\dot{A}=-i\xi e^{i\omega t}C,  \eqnum{A1a}
\end{equation}

\begin{equation}
\dot{B}=-i\xi e^{i\omega t}\left( 1+D\right) ,  \eqnum{A1b}
\end{equation}

\begin{equation}
\dot{C}=-i\Omega C-i\sum_jg(\omega _j)E_j-i\xi e^{-i\omega t},  \eqnum{A1c}
\end{equation}

\begin{equation}
\dot{D}=-i\Omega (1+D)-i\sum_jg(\omega _j)C_{j,}  \eqnum{A1d}
\end{equation}

\begin{equation}
\dot{E}_j=-i\omega _jE_j-ig(\omega _j)C,  \eqnum{A1e}
\end{equation}

\begin{equation}
\dot{C}_j=-i\omega _jC_j-ig(\omega _j)(1+D),  \eqnum{A1f}
\end{equation}
By solving eqs(A1), we get formal solutions of these time-dependent
coefficients as eqs.(9). We have introduced four new function as
\begin{equation}
u(t)={\cal L}^{-1}\left\{ \widetilde{u}[s]\right\} ,  \eqnum{A2a}
\end{equation}

\begin{equation}
u_j(t)={\cal L}^{-1}\left\{ \frac{-ig(\omega _j)\widetilde{u}[s]}{s+i(\omega
_j-\Omega )}\right\} ,  \eqnum{A2b}
\end{equation}

\begin{equation}
w(t)={\cal L}^{-1}\left\{ \frac{-i\xi \widetilde{u}[s]}{s+i\delta }\right\} ,
\eqnum{A2c}
\end{equation}

\begin{equation}
v_j(t)={\cal L}^{-1}\left\{ \frac{-ig(\omega _j)\widetilde{w}[s]}{s+i(\omega
_j-\Omega )}\right\} ,  \eqnum{A2d}
\end{equation}
where ${\cal L}^{-1}$ denotes the inverse Laplace transformation, and $%
\widetilde{u}[s]=\frac 1{s+\widetilde{K}[s]}$. Now we need to determine the
explicit form of the kernal function $K(t-t^{\prime })=\sum_j|g(\omega
_j)|^2e^{-i(\omega _j-\Omega )(t-t^{\prime })}$, and further $\widetilde{K}%
[s]$, the Laplace transformation of $K(t)$. As the standard treatment \cite
{liu2} we firstly change the sum $\sum_j$ in $K(t-t^{\prime })$ into the
integration $\frac L{\pi c}\int_0^\infty {\rm d}\omega _j$, where $L$ is the
length of the cavity and $c$ is the speed of the light in the vacuum~\cite
{rl}, i.e.,
\begin{equation}
K(t-t^{\prime })=\frac{\eta ^2\Gamma ^2NL}{\pi c}\int_0^\infty \frac{%
e^{-i(\omega _j-\Omega )(t-t^{\prime })}}{(\omega _j-\Omega )^2+\Gamma ^2}%
{\rm d}\omega _j.  \eqnum{A3}
\end{equation}
If we assume that $\Omega $ is much larger than all other quantities of the
dimension of frequency and $\Gamma $ is small quantity. Then we may adopt to
the standard approximation of extending the lower limit of the integral
Eq.(A3) to $-\infty $. By integrating eq.(A3) we get
\begin{equation}
K(t-t^{\prime })=M\Gamma e^{-\Gamma |t-t^{\prime }|},  \eqnum{A4}
\end{equation}
with $M=\frac{N\eta ^2L}c$. As long as the kernal function is determined,
one can solve time-dependent functions defined in eqs.(A2) with the help of
the Laplace transformation of $K(t-t^{\prime })$.


\begin{references}
\bibitem[a]{email}  Electronic address: suncp@itp.ac.cn
\bibitem[b]{www}  Internet www site: http:// www.itp.ac.cn/\symbol{126}suncp
\bibitem{QI} D. Bouwmeester, A. K. Ekert, and A. Zeilinger (Eds), {\it The
physics of quantum information} (Springer-Verlag, Berlin,
Heidelberg, New York, 2000).

\bibitem{zur}  W. H. Zurek, Phys.Today, {\bf 44(10)}, 36 (1991).

\bibitem{cal}  A. O. Caldeira and A. J. Leggett, {\it Ann. Phys}. (N.Y.),
{\bf 149, }374(1983); A.J. Leggett, S. Chakravarty, A.T. Dosey, M.P.A
.Fisher and W. Zwerger, Rev.Mod.Phys, {\bf 59},1-87(1987).

\bibitem{yu&sun}  L. H. Yu and C. P. Sun, Phys. Rev. A 49, 592(1994);
C.P.Sun and L. H. Yu, ibid. 51, 1845 (1995).

\bibitem{zur2}  W. H. Zurek, Philos. Trans. R. Soc. London, Ser. A 356, 1793
(1998).

\bibitem{sun}  C. P. Sun, H. Zhan and X. F. Liu, Phys. Rev. A58, 1810(1998).

\bibitem{op}  Gang. Chen, N. H. Bonadeo, D. G. Stell, D. Gammon, D. S.
Katzer, D. Park, and L. J. Sham, Science 289, 1906(2000).

\bibitem{bayer}  M. Bayer, P. Hawrylak, K. Hinzer, S. Fafard, M.
Korkusinski, Z. R. Wasilewski, O. Stern, and A. Forchel, Science 291,
451(2001).

\bibitem{quiroga} L. Quiroga, and N. F. Johnson, Phys. Rev. Lett.
{\bf 83}, 2270, (1999)
\bibitem{yi} X. X. Yi, G. R. Jin, and D. L. Zhou, Phys. Rev. A
{\bf 63}, 062307 (2001)

\bibitem{QS}  Jonathan R. Friedman, Vijav Patel, W. Chen, S. K. Tolpygo, and
J. E. Lukens, Nature 406, 43 (2000); Caspar H. van der Wal, A. C. J. ter
Haar, F. K. Wilhelm, R. N. Schouten, C.P.M. Harmans, T. P. Orlando, Seth
Lloyd, and J.E. Mooij, Science 290, 773 (2000).

\bibitem{liu2}  Yu-xi. Liu, C. P. Sun, and S. X. Yu, Phys. Rev. A63,
033816(2000).

\bibitem{louisell}  W. H. Louisell, {\it Quantum Statistical Properties of
Radiation }(John Wiley \& Sons, New York, 1990).

\bibitem{jin}  G.R. Jin, D.L. Zhou, Yu-xi Liu, X.X. Yi and C.P. Sun,
quant-ph/0101078, 2001.

\bibitem{liu1}  Yu-xi Liu, C. P. Sun, S. X. Yu, and D. L. Zhou, Phys. Rev.
A63, 023802(2000).

\bibitem{scu}  M. O. Scully, M. S. Zubairy, {\it Quantum Optics}, (Cambridge
University Press 1997).

\bibitem{Law}  C. K. Law, T. W. Chen, and P. T. Leung, Phys. Rev. A61,
023808(2000).

\bibitem{Hak}  H. Haken, Quantum field theory of solids (North-Holland
Publishing Company, 1976).

\bibitem{oliver}  M. C. de Oliveira, M. H. Moussa, and S. S. Mizrahi
Phys. Rev. A61, 63809 (2000).

\bibitem{Sun&Gao}  C. P. Sun, Y. B. Gao, H. F. Dong, and S. R. Zhao, Phys.
Rev. E 57, 3900 (1998).

\bibitem{zoller}  J. F. Poyatos, J. I. Cirac, and P. Zoller, Phys. Rev.
Lett. 77, 4728 (1996).

\bibitem{piza}  K. M. Fonseca Romero, M. C. Nemes, J. G. Peixoto de Faria,
A. N. Salgueiro, and A. F. R. de Toledo Piza, Phys. Rev. A 58, 3205 (1998).

\bibitem{c4}  M. C. de Oliveira, W. J. Munro, Phys. Rev. A61,
42309(2000).

\bibitem{rl}  Roy Lang, M. O. Scully, and Jr. Willis E. Lamb, Phys. Rev. A7,
1788(1973).
\end{references}
\end{document}